\documentclass[prl,showpacs,twocolumn,preprintnumbers,amsmath,amssymb]{revtex4}

\usepackage{amsmath} 
\usepackage{amssymb}
\usepackage{graphicx} 
\usepackage{array}
\usepackage{epstopdf}
\usepackage{textcomp}

\newcommand{\etal}{{\it et al.}}
\newcommand{\ie}{{\it i.e.}}

\newcommand{\Ca}{CaFe$_2$As$_2$ }
\newcommand{\degC}{~\textdegree C }

\begin{document}

\title{Influence of growth flux solvent on anneal-tuning of ground states in \Ca}

\author{Connor Roncaioli}
 \affiliation{Center for Nanophysics and Advanced Materials, Department of Physics, University of Maryland, College Park, MD 20742}
\author{Tyler Drye}
 \affiliation{Center for Nanophysics and Advanced Materials, Department of Physics, University of Maryland, College Park, MD 20742}
\author{Shanta Saha}
 \affiliation{Center for Nanophysics and Advanced Materials, Department of Physics, University of Maryland, College Park, MD 20742}
\author{Johnpierre Paglione}
 \email{paglione@umd.edu}
 \affiliation{Center for Nanophysics and Advanced Materials, Department of Physics, University of Maryland, College Park, MD 20742}

\date{\today}


\begin{abstract}

The effects of anneal-tuning of single-crystalline samples of \Ca synthesized via a molten Sn flux method are investigated using x-ray diffraction, chemical composition, electrical transport and magnetic susceptibility measurements in order to understand the role of growth conditions on the resultant phase diagram. Previous studies of \Ca crystals synthesized using a self-flux (FeAs) method revealed an ability to tune the structural and magnetic properties of this system by control of post-synthesis annealing conditions, resulting in an ambient pressure phase diagram that spans from tetragonal/orthorhombic antiferromagnetism to the collapsed tetragonal phase of this system. In this work, we compare previous results to those obtained on crystals synthesized via Sn flux, finding similar tunability in both self- and Sn-flux cases, but less sensitivity to annealing temperatures in the latter case, resulting in a temperature-shifted phase diagram. 

\end{abstract}
\maketitle

\section{Introduction}

In 2008, the discovery of iron-based superconductors profoundly altered the course of superconductivity research \cite{Paglione_Review,Stewart_Review,Johnston_Review}. With a superconducting transition of T$_c$ = 26~K in flourine-doped LaOFeAs,\cite{LaOFeAs} research into a broad class of Fe-pnictide materials began, including the ``122'' Fe-As materials which crystalize in the ThCh$_2$Si$_2$ structure. Within this family, the MFe$_2$As$_2$ compounds (M = Ca, Sr, Ba) have been host to particularly interesting physics. All three materials entail a tetragonal (T) $I4/mmm$ crystal structure at room temperature that undergoes an orthorhombic (O) distortion concomitant with the onset of an antiferromagnetic (AFM) order at temperatures between 135-200~K \cite{Kevin_TetraTuning}. It is widely thought that the suppression of the O/AFM ordering temperature to a quantum critical point is the origin of pairing fluctuations that drive superconductivity in the majority of iron-based superconducting materials. 

The close relationship between magnetic, superconducting and structural order has been the focus of many studies of these systems to date. However, the \Ca member of this family is one of the most unique compounds in this family, in that its unit cell dimensions lie close to a strong structural collapse instability where a dimerization of the As-As interlayer bond causes a strong symmetry-preserving change in the unit cell dimensions \cite{Hoffman}. In the iron-pnictide family, this simultaneous crystallographic and electronic structure change is of immense interest, since it appears to both suppress \cite{YuPRB} and even enhance \cite{NakajimaPRB} superconductivity. 

All members of the 122 family are susceptible to this collapse, including Sr- and Ba-based versions \cite{SrCollapse,BaCollapse}, and even the alkali metal (hole-doped) member KFe$_2$As$_2$ \cite{NakajimaPRB}, but typically requires substantial applied pressures to reach the critical interlayer As-As distance of 3~\AA\ \cite{SahaPRB,KevinSrdopedcT}.
The \Ca member lies closest to the collapsed tetragonal (cT) phase at ambient pressure, and requires only a modest amount of applied pressure \cite{Canfield_Pressure,YuPRB} or chemical pressure \cite{SahaPRB} to drive it to collapse. 
In fact, Ran \etal\ reported an interesting method of stabilizing the cT phase in \Ca at ambient pressure via annealing and quenching techniques: crystals of \Ca grown in a self (FeAs) flux with different temperature profiles were shown to exhibit a controllable variation of the O/AFM transition temperature that could be decreased by increasing post-growth annealing temperatures, and even driven into a cT phase that is stable at ambient pressure \cite{Canfield_FeAsAnn}. 
The anneal/quench tunability of \Ca was attributed a finite, temperature-dependent width of formation for \Ca that allows for a solid solubility of FeAs nanoprecipitate inclusions in single crystals.
This established that minor changes in initial growth conditions for FeAs-flux grown samples can result in drastically different sample characteristics if not post-treated with a standardized annealing, in particular inducing large lattice strain fields that mimic pressurized conditions in CaFe$_2$As$_2$.

In this study, we investigate the sensitivity of anneal-tuning of \Ca crystals to growth method initial conditions, in particular conducting a systematic study of the annealing-dependent phase diagram for Sn flux-grown crystals. We compare temperature scales in our study with those of the self-flux study of Ran \etal\ \cite{Canfield_FeAsAnn} to conclude that Sn flux-grown crystals follow the same anneal-tuning phenomenology but requiring higher temperatures, suggesting the the presence of excess FeAs during growth conditions is not a vital component for the creation of strain-induced variations in the ground state of CaFe$_2$As$_2$.

\begin{figure}[!t]
\centering
    \includegraphics[width=3.4in]{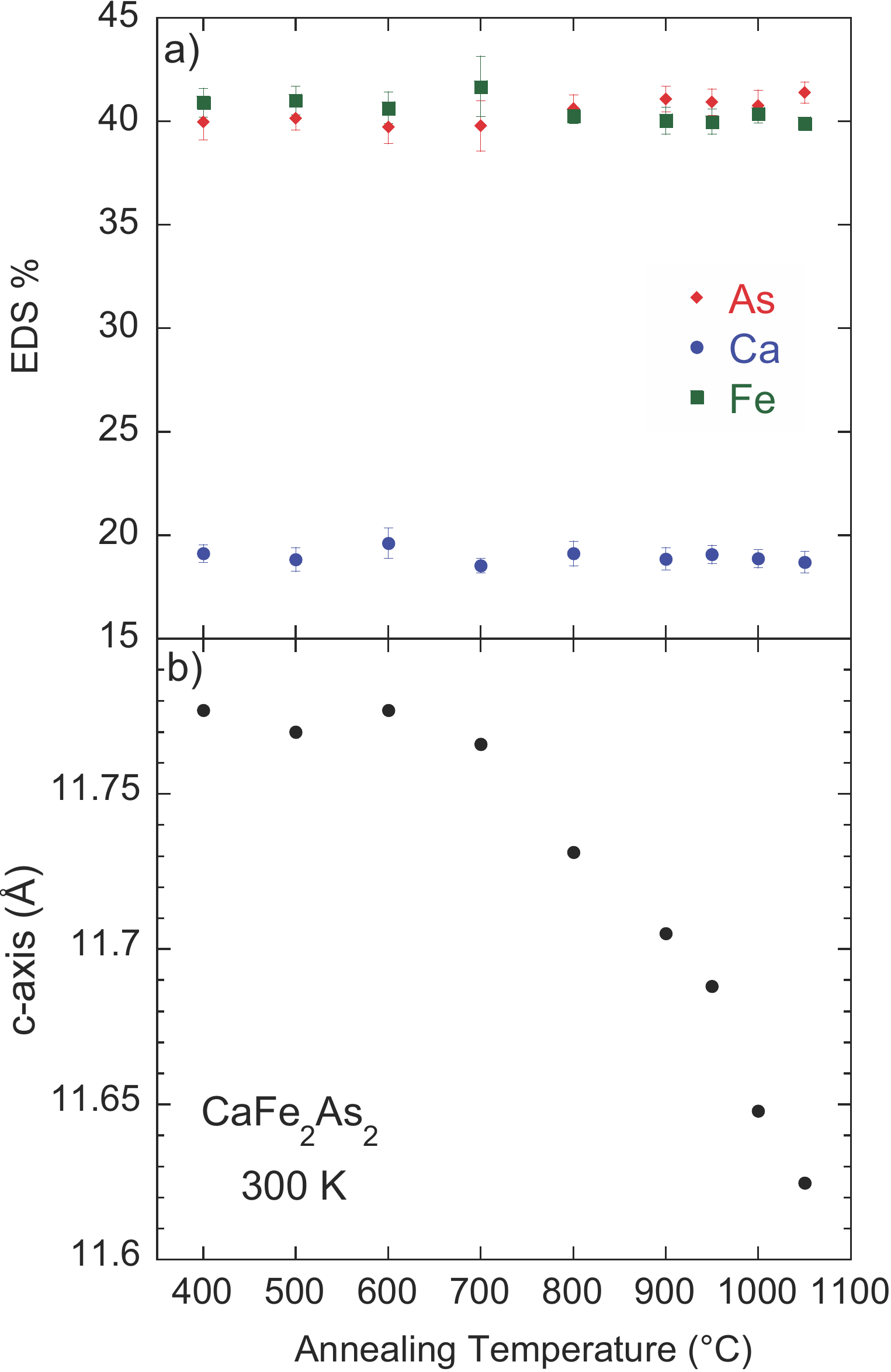}
  \caption{\label{EDS} 
  Elemental and structural characterization of \Ca as a function of annealing temperature treatment (see text). (a) Elemental analysis was determined using energy dispersive x-ray spectroscopy (EDS). All samples exhibit nominal 1:2:2 composition to within experimental error, with samples below 850\degC showing a small systematic Fe excess and samples above showing a small As excess.  (b) The $c$-axis lattice parameters were determined by x-ray diffraction from single-crystal samples, showing a notable decrease in value above 700\degC annealing temperatures.}
\end{figure}

\begin{figure}[!t]
  \includegraphics[width=3.4in]{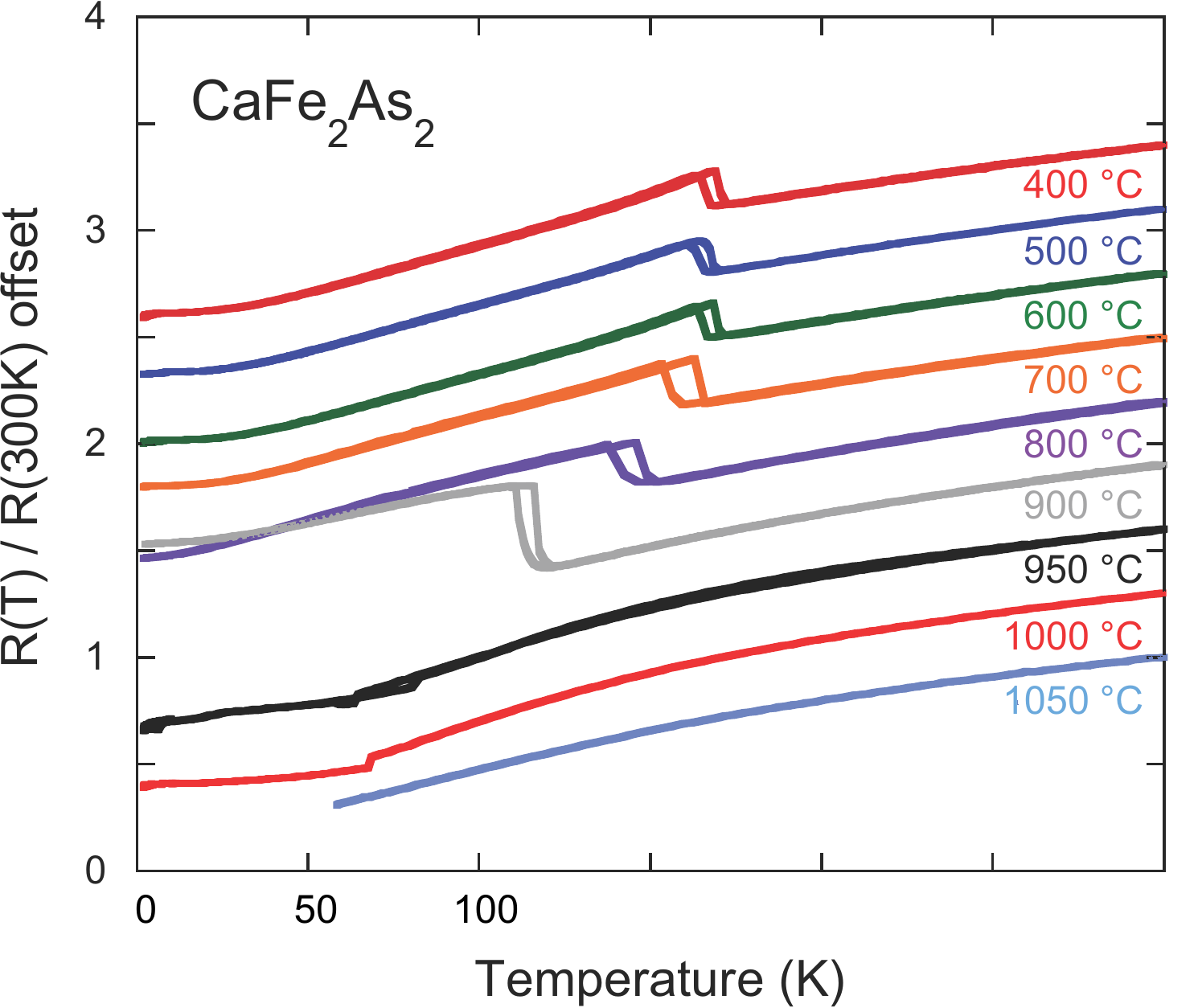}
  \caption{\label{Resistivity} 
  Effect of post-growth annealing on the electrical resistivity of Sn flux-grown crystals of \Ca (annealing temperatures shown). As the annealing temperatures approach 900\degC, the jump in resistivity found at the Tetragonal-Orthorhombic transition decreases in temperature to a position near 120~K. For higher annealing temperatures, samples show a drop and change in resistivity that corresponds to the collapsed-Tetragonal structural transition as shown in previous studies \cite{Canfield_FeAsAnn} \cite{SahaPRB}. (Data for the 1050\degC annealed sample is cut off at the collapse-Tetragonal transition due to a loss of electrical contact caused by the transition.)}
\end{figure}

\section{Experimental Methods}

Samples were grown using a molten Sn flux method in a ratio of 1:2:2:20 (Ca:Fe:As:Sn) \cite{OriginalCa}. The raw materials were collected into alumina crucibles, which do not interact with the materials during synthesis. The crucibles were then placed into quartz tubes with an inverted crucible above it with a sieving material (quartz fiber). Atmosphere was evacuated and replaced with inert Argon gas, then the tube was sealed and put in a Lindenburg box furnace. The assembly was heated to 1000~\textdegree C, slowly cooled at 2~\textdegree C/hr and then spun out at 600\degC using a centrifuge. 

This process produces large single crystals of \Ca limited in size only by the diameter of the crucible. Sample composition was verified both by x-ray and energy dispersive x-ray spectroscopy (EDS), presented in Fig. 1. Following growth, samples were annealed for one week at 400\degC to minimize and normalize temperature-dependent strain induced by the growth process, following previous work \cite{Canfield_FeAsAnn}. Samples were then annealed for 24 hours at various fixed temperatures for the purposes of this study.

Following the annealing procedure, samples were re-measured using EDS to confirm sample composition, and single-crystal x-ray to track any annealing dependent changes in the room temperature $c$-axis lattice constants. Samples retained their 1:2:2 Ca:Fe:As composition, although there was a change from slight Fe excess to slight As excess for samples annealed above 800\degC. (see Fig. 1). X-ray diffraction was performed on single-crystal samples to obtain the $c$-axis lattice parameter using the Bragg condition instead of powdered samples in order to avoid inducing any systematic strain via grinding method that could induce systematic errors in interpreting the effects of annealing. As shown in Fig.~1, the room temperature $c$-axis lattice parameter shows a sensitivity to annealing temperature, decreasing at higher temperatures by as much as 0.15~\AA, a $\sim$1\% change in lattice parameter. 

Electrical resistivity $\rho(T)$ of annealed samples was measured using the four-wire technique in a commercial cryostat. Samples annealed at 400\degC exhibit a sharp hysteretic jump in resistivity at 160~K that corresponds to the tetragonal-orthorhombic (TO) transition in \Ca materials \cite{Canfield_FeAsAnn}. As annealing temperature is increased, the TO resistive transition drops in temperature but steadily grows in magnitude until at 950\degC the upwards jump disappears and is replaced by a drop in resistivity. This corresponds to the first-order transition into the non-magnetic cT ground state \cite{Canfield_FeAsAnn,SahaPRB}, which involves an abrupt 10\% shift in the $c$-axis, and can have damaging effects on the sample and contacts as it undergoes such a large volume change. (In the case of the 1050\degC sample, the cT transition caused our four-wire contacts and thus resistivity measurement to fail, which is removed from the figure for clarity.) The annealing effect is completely reversible as shown in Fig.~3, where samples subjected to various annealing and quenching conditions exhibit the entire range of behavior. For example, samples annealed at 1000\degC for 24 hours followed by a fast ambient air quench (\ie, removed from furnace) are left in the cT phase regime, while following the same 1000\degC anneal by a slower furnace cool (\ie, shut off furnace and let cool over $\sim$24 hours) as well as a second seven-day anneal at 400\degC recovers the original intrinsic (ie, non-strained) condition with TO transition at its maximum of 165~K.

Magnetic susceptibility $\chi(T)$ was measured using post-annealed \Ca samples spanning the range of ground states. As shown in Fig.~5, this includes samples exhibiting TO transitions (700 and 900\degC) as well as cT transitions (950 and 1000\degC), which all show drops in $\chi(T)$ that match the features found in the resistivity measurements. While both types of transition cause a similar abrupt drop in $\chi(T)$ at the transition, it is of note that the 900\degC annealed sample exhibits a broadened TO transiition as expected by the induced strain, while the higher temperature annealed sample transitions are much sharper and do not broaden close to or far away from the transition. This is due to the first-order nature of the structural transition to the cT state, which occurs very abruptly. In addition, as shown in Fig.~5 the higher temperature annealed samples exhibit a low-temperature hump in magnetization below the cT transitions, whose origin remains to be determined.

\begin{figure}[!t]
  \includegraphics[width=3.4in]{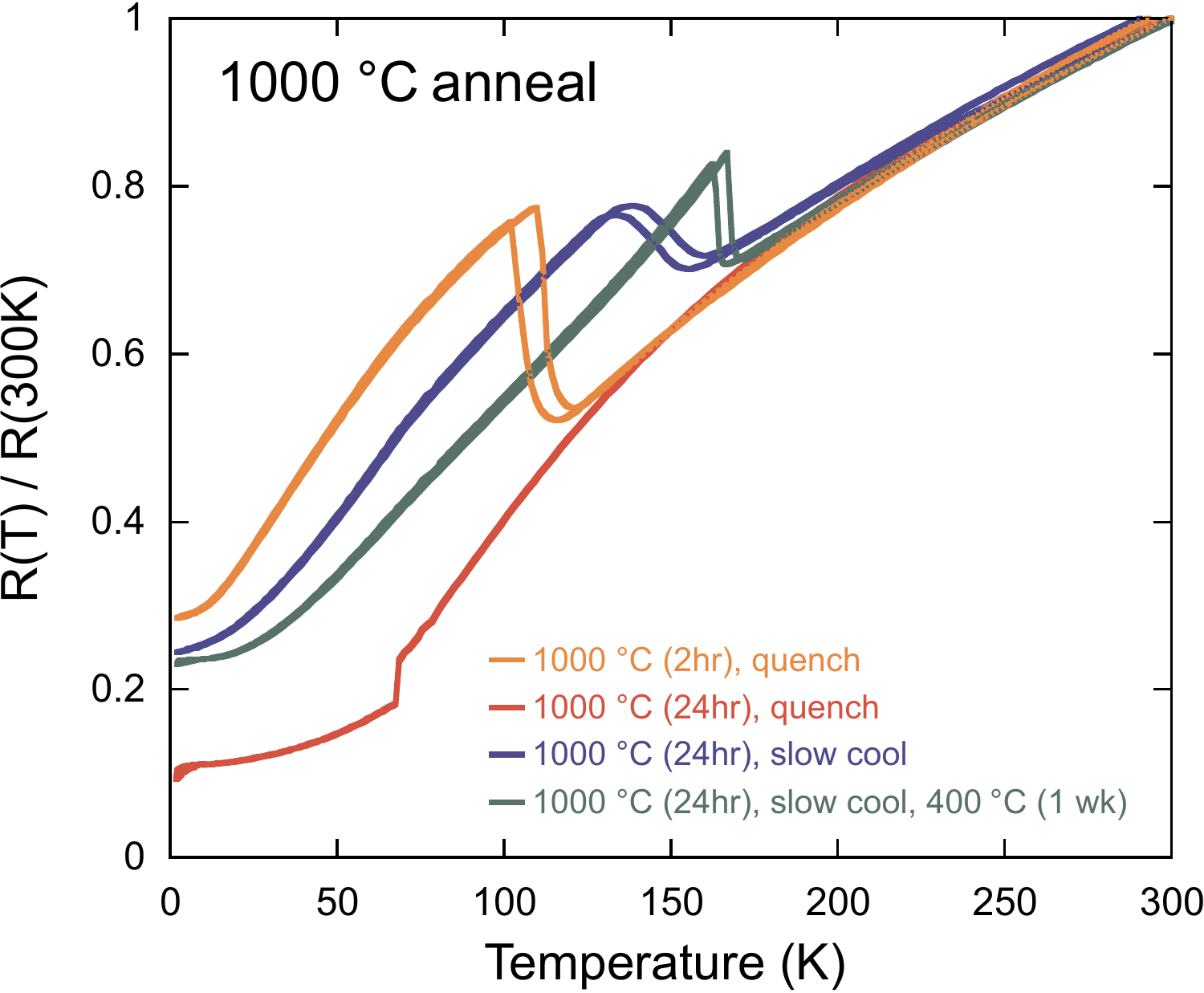}
  \caption{\label{Temperature Variance} 
  Systematic study of different annealing conditions on crystals of \Ca, where all samples were annealed at 1000\degC (after standard 400\degC anneal - see text) for either 2 or 24 hours as indicated, followed by further processing involving either quench (sealed ampoule removed from furnace and quickly cooled in ambient air), slow cool (furnace power shut off and sample slowly cooled over $sim$24 hours), and subsequent annealing at 400\degC in one case.
  As shown, the varying treatments produce the entire range of states in \Ca, including partial suppression of the tetragonal-orthorhombic transition (1000~\textdegree C/2hr), full tuning into the collapsed tetragonal state (1000~\textdegree C/24hr), and recovery of nearly (slow cool) and fully (slow cool + 400~\textdegree C/1 week) restored strain-free state as indicated by a maximized 165~K tetragonal-orthorhombic transition.} 
\end{figure}

\begin{figure}[!t]
  \includegraphics[width=3.4in]{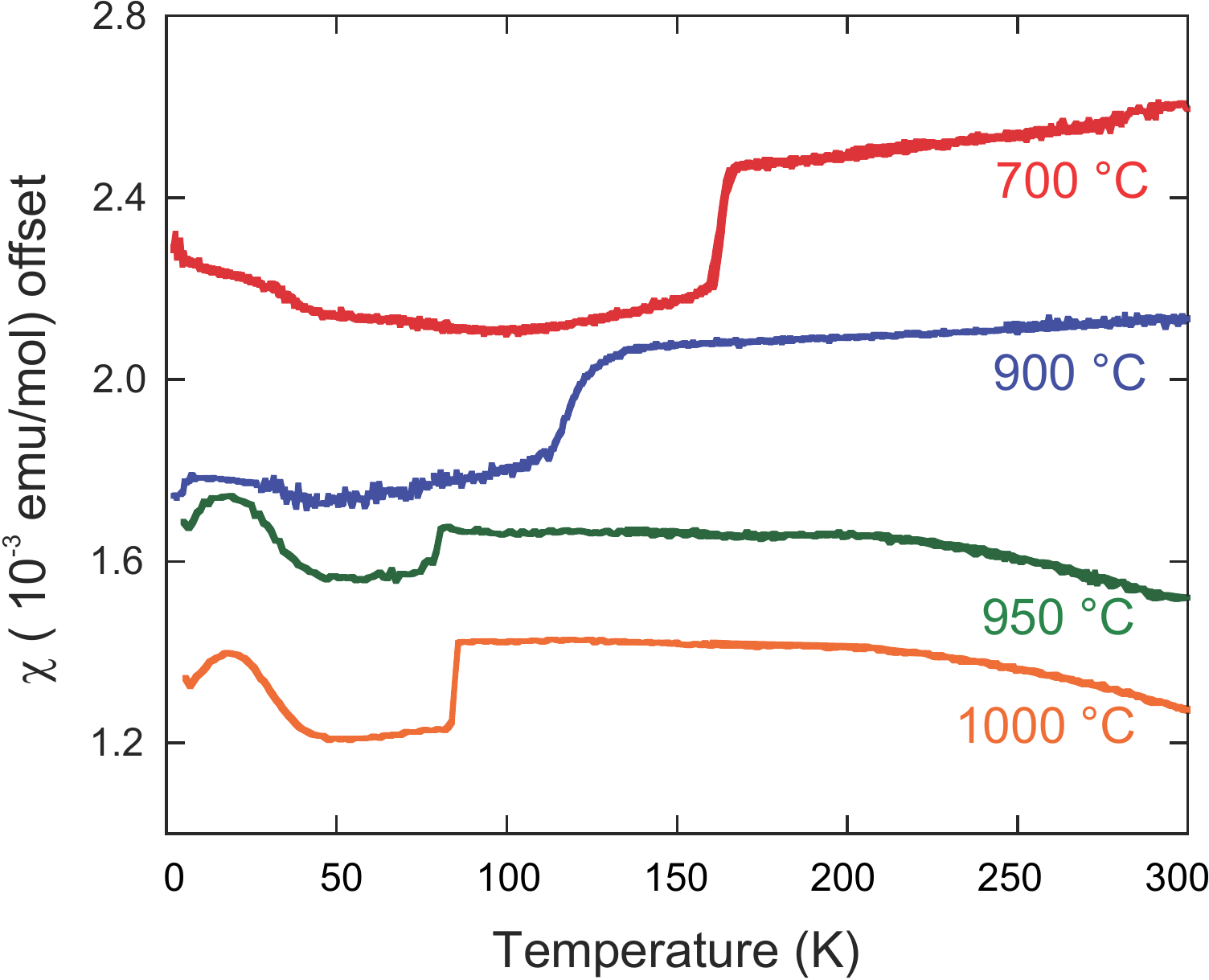}
  \caption{\label{Magnetization}
 Magnetic susceptibility of annealed crystals of \Ca spanning the range of anneal-tuned ground states. The 700\degC and 900\degC samples show slightly broadened transitions as expected for the transition from high temperature tetragonal to low temperature orthogonal ground state. Similarly, the 950\degC and 1000\degC samples show a decrease in susceptibility associated with the collapse transition for these samples, which appears fairly sharp as expect for such an abrupt volume collapse.}
\end{figure}

\section{Discussion}

The previous study of Ran \etal, which studied the response of FeAs flux-grown \Ca crystals to post-growth annealing, found a particular sensitivity to annealing treatments due to interstitial FeAs nanoprecipitates \cite{Canfield_FeAsAnn}. In this case, higher temperature synthesis or annealing conditions are thought to spread interstitial inclusions of binary FeAs throughout the lattice, creating a strain field responsible for causing the collapse transition to occur at ambient pressure. This picture was confirmed by transition electron microscopy measurements that found a tweed-like structure parallel to the \{h,0,0\} plane in strained materials, and condensed precipitates in samples that had been annealed at low temperatures to reduce the strain. 

As shown in Fig.~5, the results of our study of Sn flux-grown \Ca are quite comparable to the previous work, exhibiting a controllable tuning of structural ground states and transitions via post-annealing treatment. Our samples exhibit transitions from high-temperature tetragonal to low-temperature orthorhombic for annealing temperatures from 400\degC (\ie, baseline anneal) to at least 900~\textdegree C, with a suppression of the TO transition temperature as well as a shortening of the tetragonal $c$-axis with increasing annealing temperature as observed by Ran \etal. At higher annealing temperatures, the induced tetragonal to cT transition is also present as expected by previous work. While the overall trend and resultant phase diagram is nearly identical in both cases, the TO transition regime spans a wider annealing temperature range for the Sn flux case, reaching $\sim$200\degC higher in tuning temperature as shown in Fig.~5, requiring higher temperatures to induce the cT phase.

This suggests that the mechanism of strain is reduced, though not eliminated, in Sn-flux grown CaFe$_2$As$_2$. Such a conclusion is a reasonable expectation, given that the flux solvent spin-off temperature is a full 400\textdegree\ lower in the Sn case, providing a much larger temperature window for more effective annealing of crystals in the Sn melt after formation at +1000\degC. This wider window may alter the width of formation such that the resultant crystals are less susceptible to strain field formation via anneal-tuning. However, the simple fact that anneal-tuning is still an effective means of controlling strain in \Ca suggests that the situation is not so simple. Another aspect of the picture involves understanding the role of Sn inclusions in the Sn flux-grown crystals. Because it is not uncommon to observe a trace of superconductivity due to elemental Sn in these samples, it is possible that inclusions of strain-relieving Sn-based pockets occur in Sn-grown samples similar to the FeAs precipitates in FeAs-grown \Ca. In both cases,  the critical (300~K) $c$-axis dimension of ~11.7~\AA \ appears to be the same value required to induce the low-temperature cT state, suggesting similar strain environments tuned by slightly different energy scales. While the question of whether annealing-dependent behavior in \Ca is intrinsic to the lattice or partially dependent on the existence of nanoprecipitates or inclusions remains open, it is clear that the role of growth conditions has a moderately strong effect on this phenomenon.

\begin{figure}[!t]
  \includegraphics[width=3.4in]{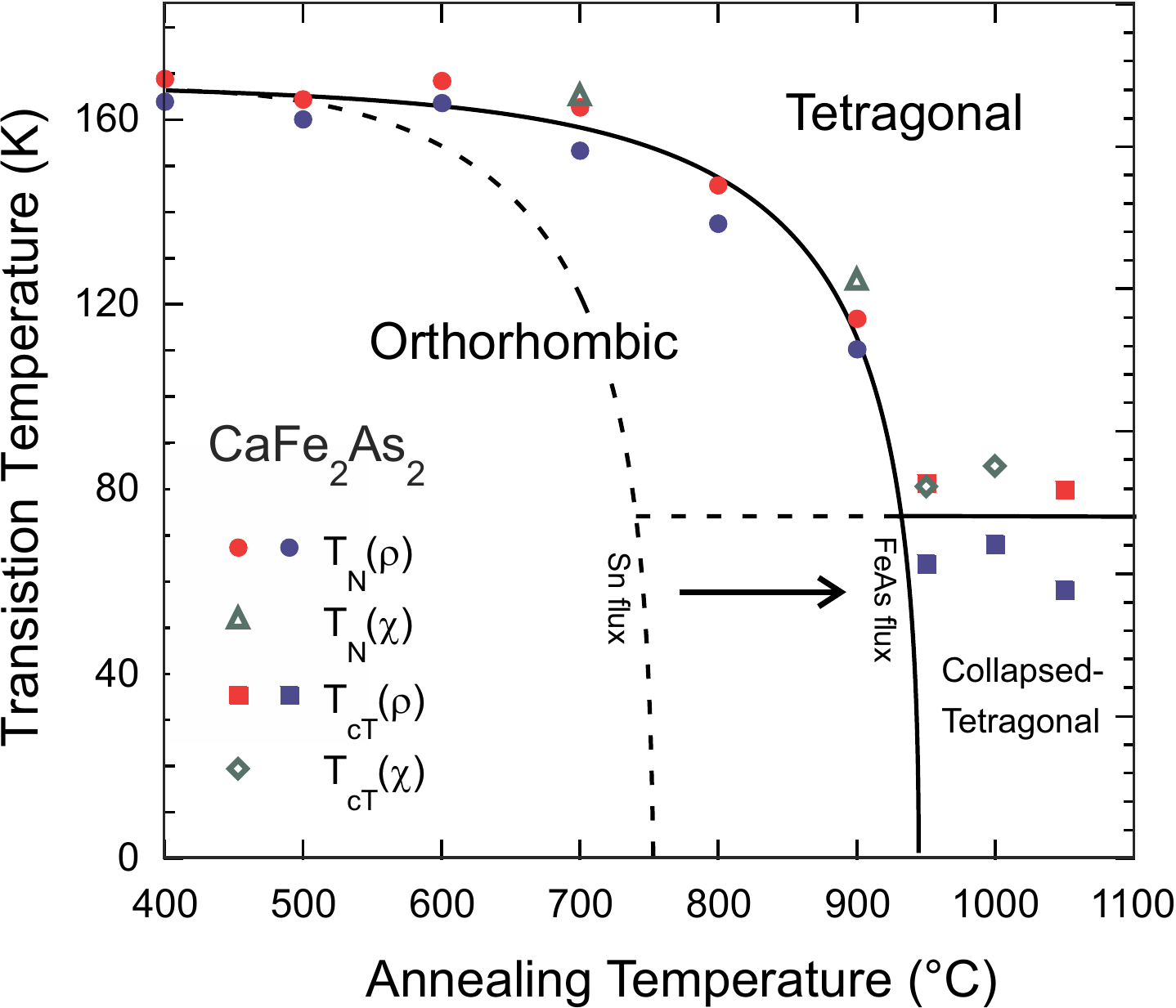}
  \caption{\label{Phase Diagram} 
 Phase diagram combining resistivity (Fig. 2) and magnetic susceptibility (Fig. 4) transition temperature data (solid lines), plotted as a function of applied annealing temperature for Sn flux-grown samples of \Ca. The transition between high temperature tetragonal and low-temperature orthorhombic phases is shown to decrease with increasing annealing temperatures, falling quickly near 950\degC to give way to a transition to the collapsed tetragonal phase at higher annealing temperatures.
This quite closely mirrors the behavior of the anneal tuning of FeAs-flux grown \Ca as determined by Ran \etal (dashed lines) \cite{Canfield_FeAsAnn}, but with phase boundaries shifted by approximately +200\degC as shown by the arrow.} 
\end{figure}

\section{Conclusion}

In this work, we studied the effects of post-synthesis annealing on Sn flux-grown crystals of CaFe$_2$As$_2$. The results indicate that the ambient condition structural state of \Ca can be anneal-tuned by heat treatments such that the high-temperature tetragonal to low-temperature orthorhombic structural transition, which occurs at 165~K in the nominally strain-free CaFe$_2$As$_2$, can be suppressed with increasing annealing temperatures. Beyond approximately 950\degC, the tetragonal-orthorhombic transition is fully suppressed and is replaced by a collapsed tetragonal ground state. The evolution of $c$-axis lattice parameters and resultant structural states with anneal-tuned strain is very close to that found in \Ca samples synthesized using a FeAs flux as studied previously by Ran \etal\ \cite{Canfield_FeAsAnn} but with less responsiveness to annealing temperatures, suggesting a moderately strong sensitivity of CaFe$_2$As$_2$ ground states to thermally induced strain even without excess FeAs present during flux synthesis.

%




\begin{thebibliography}{10}



\bibitem{Paglione_Review}
J. Paglione, R. Greene, Nature Physics, {\bf 6}, 645-658 (2010)

\bibitem{Stewart_Review}
G.R. Stewart, Rev. Mod. Phys., {\bf 83}, 1589-1652 (2011)

\bibitem{Johnston_Review}
D. C. Johnston, Adv. Phys., 59:6, 805-1061, (2010)

\bibitem{LaOFeAs}
H. Takahashi, K. Igawa, K. Arii, Y. Kamihara, M. Hirano, H. Hosono, Nature, {\bf 453}, 376-378 (2008) 

\bibitem{Kevin_TetraTuning}
K. Kirshenbaum, N.~P. Butch, S.~R. Saha, P.~Y. Zavalij, B.~G. Ueland, J.~W. Lynn, J. Paglione, Phys. Rev. B, {\bf 86}, 060504 (2012)

\bibitem{Hoffman}
I. Zeljkovic, D. Huang, C.~L. Song, B. Lv, C.~W. Chu, J.~E. Hoffman, Phys. Rev. B, {\bf 87}, 201108 (2013)

\bibitem{YuPRB}
W. Yu, A.~A. Aczel, T.~J. Williams, S.~L. Bud'ko, N. Ni, P.~C. Canfield, G.~M. Luke, Phys. Rev. B, {\bf 79}, 020511 (2009)

\bibitem{NakajimaPRB}
Y. Nakajima, R. Wang, T. Metz, X. Wang, L. Wang, H. Cynn, S.~T. Weir, J. Paglione, Phys. Rev. B, {\bf 91}, 060508 (2015)

\bibitem{SrCollapse}
W.~O. Uhoya, J.~M. Montgomery, G.~M. Tsoi, Y.~K. Vohra, M.~A. McGuire, A.~S. Sefat, B.~C. Sales, S.~T. Weir, J. Phys. Condens. Mat., {\bf 23}, 1232201 (2011)

\bibitem{BaCollapse}
R. Mittal, {\it et al.}, Phys. Rev. B, {\bf 83}, 054503 (2011)

\bibitem{SahaPRB}
S.~R. Saha, N.~P. Butch, T. Drye, J. Magill, S. Ziemak, K. Kirshenbaum, P.~Y. Zavalij, J.~W. Lynn, J. Paglione, Phys. Rev. B, {\bf 85}, 024525 (2012)

\bibitem{KevinSrdopedcT}
J.~R. Jeffries, {\it et al.}, Phys. Rev. B, {\bf 90}, 144506 (2014)

\bibitem{Canfield_Pressure}
M.~S. Torikachvili, S.~L. Bud'ko, N. Ni, P.~C. Canfield, S.~T. Hannahs, Phys. Rev. B, {\bf 80}, 014521 (2009)

\bibitem{Canfield_FeAsAnn}
S. Ran, {\it et al.}, Phys. Rev. B, {\bf 83}, 144517 (2011) 

\bibitem{OriginalCa}
N. Ni, S. Nandi, A. Kreyssig, A.~I. Goldman, E.~D. Mun, S.~L. Bud'ko, P.~C. Canfield, Phys. Rev. B, {\bf 78}, 014523 (2008)

\bibitem{cTSC}
P.~C. Canfield, {\it et al.}, Physica C, {\bf 469}, 404-412 (2009)

\bibitem{CanfieldNMR}
Y. Furukawa, B. Roy, S. Ran, S.~L. Bud'ko, P.~C. Canfield, Phys. Rev. B, {\bf 89}, 121109 (2014)

\bibitem{Chen}
J.~J. Ying, {\it et al.}, Phys. Rev. Lett., {\bf 107}, 067001 (2011)

\bibitem{NeutronhighTcs}
J.~W. Lynn, P. Dai, Physica C, {\bf 469} 469-476 (2009)

\bibitem{cTtuning}
K. Zhao, C. Stingl, R.~S. Manna, C.~Q. Jin, P. Gegenwart, Phys. Rev. B, {\bf 92} 235132 (2015)

\bibitem{Yoshizawa}
S. Simayi, {\it et al.}, J. Phys. Soc. Jpn., {\bf 82} 114604 (2013)

\bibitem{DefectStates}
M.~N. Gastiasoro, I. Paul, Y. Wang, P.~J. Hirschfeld, B.~M. Andersen, Phys. Rev. Lett., {\bf 113} 127001 (2014)

\bibitem{Feng}
D.~F. Xu, {\it et al.}, Phys. Rev. B, {\bf 90}, 214519 (2014)

\bibitem{LaOFeP}
Y. Kamihara, H. Hiramatsu, M. Hirano, R. Kawamura, H. Yanagi, T. Kamiya, H. Hosono, J. Am. Chem. Soc., {\bf 128} 10012-10013

\bibitem{MultiOrbital}
O.~K. Andersen, L. Boeri, Ann. Phys., {\bf 523} 8-50 (2011)


\end{thebibliography}

\end{document}